# Blockchain in Internet of Things: Challenges and Solutions

Ali Dorri, Salil S. Kanhere, and Raja Jurdak

**Abstract**

The Internet of Things (IoT) is experiencing exponential growth in research and industry, but it still suffers from privacy and security vulnerabilities. Conventional security and privacy approaches tend to be inapplicable for IoT, mainly due to its decentralized topology and the resource-constraints of the majority of its devices. BlockChain (BC) that underpin the crypto-currency Bitcoin have been recently used to provide security and privacy in peer-to-peer networks with similar topologies to IoT. However, BCs are computationally expensive and involve high bandwidth overhead and delays, which are not suitable for IoT devices. This position paper proposes a new secure, private, and lightweight architecture for IoT, based on BC technology that eliminates the overhead of BC while maintaining most of its security and privacy benefits. The described method is investigated on a smart home application as a representative case study for broader IoT applications. The proposed architecture is hierarchical, and consists of smart homes, an overlay network and cloud storages coordinating data transactions with BC to provide privacy and security. Our design uses different types of BC's depending on where in the network hierarchy a transaction occurs, and uses distributed trust methods to ensure a decentralized topology. Qualitative evaluation of the architecture under common threat models highlights its effectiveness in providing security and privacy for IoT applications.

**Introduction**

The Internet of Things (IoT) represents one of the most significant disruptive technologies of this century. It is a natural evolution of the Internet (of computers) to *embedded and cyber-physical systems*, "things" that, while not obviously computers themselves, nevertheless have computers inside them. With a network of cheap sensors and interconnected things, information collection on our world and environment can be achieved at a much higher granularity. Indeed, such detailed knowledge will improve efficiencies and deliver advanced services in a wide range of application domains including pervasive healthcare and smart city services. However, the increasingly invisible, dense and pervasive collection, processing and dissemination of data in the midst of people's private lives gives rise to serious security and privacy concerns [1]. On the one hand, this data can be used to offer a range of sophisticated and personalized services that provide utility to the users. On the other hand, embedded in this data is information that can be used to algorithmically construct a virtual biography of our activities, revealing private behavior and lifestyle patterns.

The privacy risks of IoT are exacerbated by the lack of fundamental security safeguards in many of the first generation IoT products on the market. Numerous security vulnerabilities have been identified in connected devices ranging from smart locks [2] to vehicles [3]. Several intrinsic features of IoT amplify its security and privacy challenges including: lack of central control, heterogeneity in device resources, multiple attack surfaces, context-aware and situational nature of risks, and scale.

Naturally, security and privacy for IoT is receiving a lot of attention within the research community. In [4], a distributed capability-based access control method is proposed to control access to sensitive information. However, their proposed method introduces excessive delays and overheads and could potentially compromise user privacy. Authors in [5] used IPsec and TLS to provide authentication and privacy, but these methods are computationally

*Ali Dorri and Salil S. Kanhere are with The University of New South Wales (UNSW); Raja Jurdak is with CSIRO Brisbane.*

expensive and may thus be inappropriate for many resource-limited IoT devices. A privacy management method is proposed in [6] which measures the risk of disclosing data to others, however, in many circumstances, the perceived benefit of IoT services outweigh the risk of privacy loss. There is thus a need for privacy-aware sharing of IoT data without sacrificing the privacy of users. In summary, these and several other prior works have yet to address the aforementioned challenges in ensuring security and privacy for IoT in a comprehensive manner.

In this article, we argue that the answer may lie in the fundamental technology that underscores emerging cryptocurrencies. Bitcoin [7], the world's first decentralized digital currency was launched in 2008. Bitcoin is underpinned by a peer-to-peer computer network that is made of its users' machines, similar to BitTorrent. In addition, a changeable Public Key (PK) is used as user's identity[1] to provide anonymity and privacy. The main technology behind Bitcoin is called BlockChain (BC), an immutable public record of data secured by a network of peer-to-peer participants. BC is rapidly gaining popularity and is being used for many other applications including smart contracts, distributed cloud storage and digital assets. BC consists of blocks chained together as a ledger. Any node in the peer-to-peer network can choose to be a miner, an entity that is responsible for mining blocks to BC by solving a resource-intensive cryptographic puzzle called Proof Of Work (POW) [8] and appending new blocks to BC. When a new transaction occurs, it is broadcast to the entire network. All miners who receive the transaction verify it by validating the signatures contained within the transaction. Each miner appends the verified transaction to its own pending block of transactions that are waiting to be mined. The robustness of the BC is ensured by the fact that multiple miners process a single transaction. However, robustness comes at a price as multiple miners have to expend their resources for mining the same transaction, which in turn also increases the delay. The following salient features of BC make it an attractive technology for addressing the aforementioned security and privacy challenges in IoT:

- Decentralization: The lack of central control ensures scalability and robustness by using resources of all participating nodes and eliminating many-to-one traffic flows, which in turn decreases delay and overcomes the problem of a single point of failure.
- Anonymity: The inherent anonymity afforded is well-suited for most IoT use cases where the identity of the users must be kept private.
- Security: BC realizes a secure network over untrusted parties which is desirable in IoT with numerous and heterogeneous devices.

However, adopting BC in IoT is not straightforward and will require addressing the following critical challenges:

- Mining is particularly computationally intensive, while the majority of IoT devices are resource restricted.
- Mining of blocks is time consuming while in most IoT applications low latency is desirable.
- BC scales poorly as the number of nodes in the network increases. IoT networks are expected to contain a large number of nodes.
- The underlying BC protocols create significant overhead traffic, which may be undesirable for certain bandwidth-limited IoT devices.

The main contribution of this position paper is to introduce a blockchain-based architecture for IoT that delivers lightweight and decentralized security and privacy. The architecture retains the benefits of BC while overcoming the aforementioned challenges in integrating BC in IoT. To exemplify our ideas, we use an illustrative example of a smart home in the rest of the article. However, our proposed architecture is application-agnostic and well-suited for diverse IoT use cases.

---

[1] This article assumes that the reader is familiar with the basic concepts of bitcoin and cryptography. Interested readers should refer to [7] for introductory resource.

**Block-based IoT Architecture**

We consider a typical smart home setting where a user, Alice has equipped her home with a number of IoT devices including a smart thermostat, smart bulbs, an IP camera and several other sensors. The proposed architecture shown in Figure 1 includes three tiers, namely the smart home (or more generally the local network), the overlay network, and the cloud storage.

We consider data store and access use cases: Alice should be able to access the data from her smart home, e.g., the current temperature in her bedroom, remotely. Moreover, smart devices should be able to store data on storages to be used by a third party (e.g., the smart thermostat provider) to avail of some services.

Prior to discussing the details of the proposed architecture, we briefly introduce the network tiers:

**Smart Home:** The smart home is comprised of the following three parts:
- **Devices:** All smart devices located in the home.
- **Local BC:** A secure and private BC that is mined and stored by one (or more) resource-capable device(s), which is always online. An example could be a smart hub or home computer. Unlike the Bitcoin BC whose management is decentralized, the local BC is centrally managed by its owner. All transactions pertaining to a particular device are chained together. The owner is responsible for adding new devices by creating a starting transaction, which is similar to creating a new coin in Bitcoin. The owner can also remove an existing device by deleting its ledger. The local BC has a policy header, which is an access control list that allows the owner to control all transactions happening in her home. Devices can communicate with each other only if the owner permits them to do so by giving them a shared key based on the generalized Diffie-Hellman algorithm [9]. While all blocks in BC have a policy header, the most updated one, placed in the header of the last block, is used for checking and changing policies. As in Bitcoin, transactions are grouped together and mined in units of blocks. However, unlike Bitcoin, each block is mined and appended to BC without POW or other puzzles to reduce the associated overheads. The miner adds a pointer to the previous block, copies the policy in the previous block header to the new block and chains the block to the BC. Another difference with Bitcoin is that when a transaction has been added to a block, it is treated as a true transaction, whether the block is mined or not.

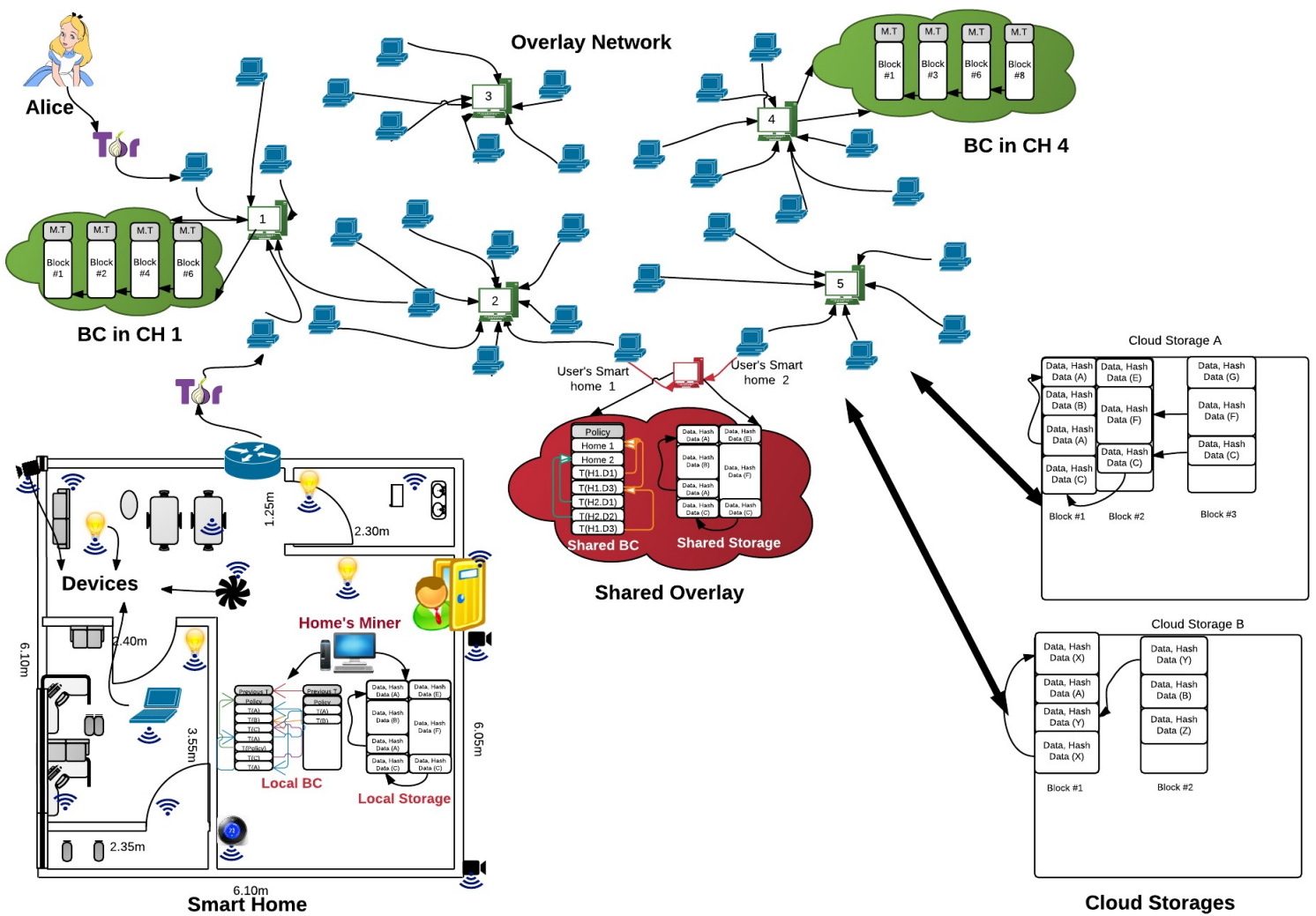

Figure 1. Overview of proposed architecture

- **Local Storage:** In each home, there may be an optional local storage for storing data locally as depicted in the smart home in Fig.1. This could be a local backup drive.

In addition to these parts, each home's miner has a list of PKs used for giving others permission to access the smart home data.

**Overlay Network**

The overlay network is akin to the peer-to-peer network in Bitcoin. The constituent nodes could be smart home miners, other high resource devices in the home, or the user's smartphone or personal computer. Each node uses Tor [10] to connect to overlay network for additional anonymity at IP-layer. A particular user may have more than one node in the overlay network. To decrease network overhead and delay, nodes in the overlay network are grouped in clusters and each cluster elects a Cluster Head (CH). Each node is free to change its cluster if it experiences excessive delays. Moreover, nodes in the cluster can elect a new CH at any time. Each CH maintains the following three lists:

- PK of requesters: the list of PKs that are allowed to access data for the smart homes connected to this cluster.
- PK of requestees: the list of PKs of smart homes connected to this cluster that are allowed to be accessed.
- Forward list: a list of transactions sent for other CHs in the network.

An overlay BC is kept by all CHs in the overlay network, which contains multisig transactions sent by the cloud storage and access transactions. Unlike Bitcoin mining, each CH independently decides whether to keep a new block or discard it, based on its communication with the received transaction's participants. This can lead to different versions of BC in each CH. For instance, in Fig 1, CH1 has blocks 1, 2, 4 and 6 while CH4 has 1, 3, 6 and 8. Since there is no requirement for the BCs to be reconciled the synchronization overheads are reduced. However, in some cases discovering a particular block or transaction comes at the cost of higher delay. In case a user has more than one home and wishes to manage them together, a shared overlay consisting of the high resource devices in the multiple homes can be formed as shown by the red cloud in Fig 1. A common miner and shared storage are selected for this shared overlay. In the overlay BC each device has a starting transaction chained to its home's starting transaction. This leads to forking in shared BC, a deviation from the Bitcoin BC where forking is not permitted due to its double spending affect. When a shared overlay exists, the high resource devices of the constituent homes maintain a table containing the block-number and hash of data for the last transaction.

**Cloud Storage**

In some cases, devices in the smart home (e.g. a smart thermostat) may wish to store their data in the cloud storage, so that a third party Service Provider (SP) can access the stored data and provide certain smart services (e.g. intelligent temperature adjustment). The cloud storage groups user's data in identical blocks associated with a unique block-number. Block-number and hash of stored data are used by the user for authentication. If the storage can successfully locate data with given block-number and hash, then the user is authenticated. Received data packets from users are stored in a First-In-First-Out order in blocks along with the hash of stored data as shown in the bottom right of Fig 1. After storing data, the new block-number is encrypted using a shared key derived from generalized Diffie-Hellman algorithm. This ensures that whoever possesses the key is the only one who knows the block-number. Since hashes are collision-resistant and only the true user knows the block-number, we can guarantee that no one other than the true user can access her data and also chain fresh data to an existing ledger. It is worth noting that each user can either create different ledgers of data in storage for each of its devices or a single common ledger for all of its devices. The former is particularly useful if the user wishes to provide access to all data of a particular device to a SP.

**Transaction Handling**

Having discussed the general topology of our BC-based IoT security and privacy architecture, we now focus on how transactions are handled.

**Storing**

Based on the defined policy, each device may store data in local, shared or cloud storage. As an example, a smart thermostat, typically stores data in the cloud storage to be used by the SP to implement certain smart services. Let's assume that Alice has created an account in a cloud storage facility and set up permissions for her thermostat to upload data to this facility. During the bootstrapping process, the cloud storage returns a pointer to the first block of data. When the smart thermostat needs to store data in the cloud storage, it sends its data to the miner. After checking permissions and extracting the previous block-number and hash, the miner creates a random ID and sends data to the storage with this ID, as shown in Fig 2a. It is assumed that at any given time, two nodes cannot have the same ID. The storage checks the validity of the transaction and also confirms that there is space available in the cloud storage. If so, it calculates a hash of received data packets and compares it with the received hash. If the two hashes match, then data packets are stored in the storage and the new block-number is encrypted with the shared key, and sent to the miner. Next, the signed hash of data is signed by the storage and sent to the overlay network to be mined in the overlay BC. This ensures that any changes in the user's data are visible to all.

Recall that shared storage is a local storage managed by the owner of the homes and is trusted. Therefore, there is no accounting and subsequently there is no need for the miner to send the hash of current data during a store operation. Also, the storage is not required to send hash of data to the overlay network. All other processes are the same as with cloud storage.

In case of local storage, the steps are similar, the only difference being that there is no need to use IDs since all communications are performed locally in the smart home.

**Accessing**

The SP may need to access the stored data for a certain time period (e.g. the past 24 hours) or the entire chain of data for a particular device, in order to implement certain services. To access information, the SP creates and signs a multisig transaction, which needs to be signed by the requester (SP) and the requestee (smart home's miner) and sent it to its own CH. The CH checks both lists of PKs. If either the multisig transaction's requester is in CH requester's PK list or it's requestee is in its requestee's PK list, then it broadcasts the transaction to its own cluster. Otherwise, the transaction is broadcasted to other CHs and the PK of requester is put in forward list. When the smart home's miner receives a multisig transaction, it has to check the policy in her local BC to verify if SP has permission to access data, which should have been granted previously by the user. If so, the miner requests packets from the storage, encrypts them with requester's PK, and sends them to the requester, as shown in Fig 2b. Prior to sending data, the miner may use methods such as safe answer or introduce noise to provide additional privacy [11]. The output of multisig transaction can be set to either '1' or '0' by the miner, indicating whether the requester has access to the data. After sending data for the requester, the miner should store multisig transaction in the local BC. In addition, the miner sends the multisig transaction to a random set of CHs to be stored in the overlay network. These stored multisig transactions can be treated as a proof that the data was sent by the user and can also be used to make other nodes aware of misbehavior (e.g. a node requests for data that it is not permitted to access). The miner may decide not to send multisig transaction to the overlay BC when it has no intention to reveal this access for others. This increases user's privacy by preventing an attacker to link different transactions to a real world identity. In a smart home, there are several instances when the owner of the home or SP needs access to the entire data of a device. To reduce the network overhead in these cases, different policy levels are defined as follow:

- If the requester is the user or an SP which is authorized to access the entire chain of data, then the miner sends block-number and hash of data in storage.
- Otherwise, the miner sends the minimum possible data that can satisfy the requester query by using methods like adding noise or safe answer.

In our system, all CHs that have forwarded a transaction have to keep it in their BC. Moreover, CHs of requester and requestee also record the transaction. Other nodes decide to store a transaction based on whether they are involved in any intermediate communication related to that transaction.

**Monitoring**

In some instances, the smart home owner may wish to access certain information from their smart home device in real-time, for example, check the current configuration of their smart thermostat. We introduce a monitor transaction for this purpose (see Fig. 2c), in which the miner requests real-time data from the requested device and sends it to the requester. This data could be sent conterminously (e.g. live camera being viewed by the user).

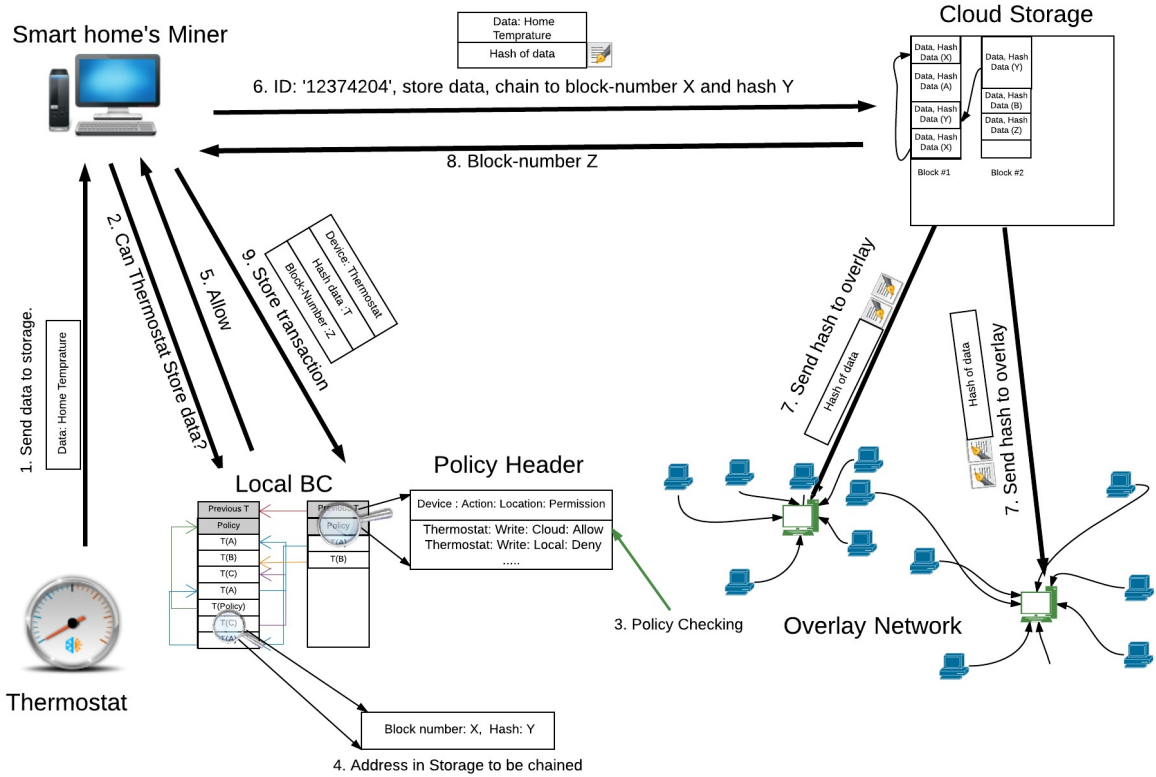

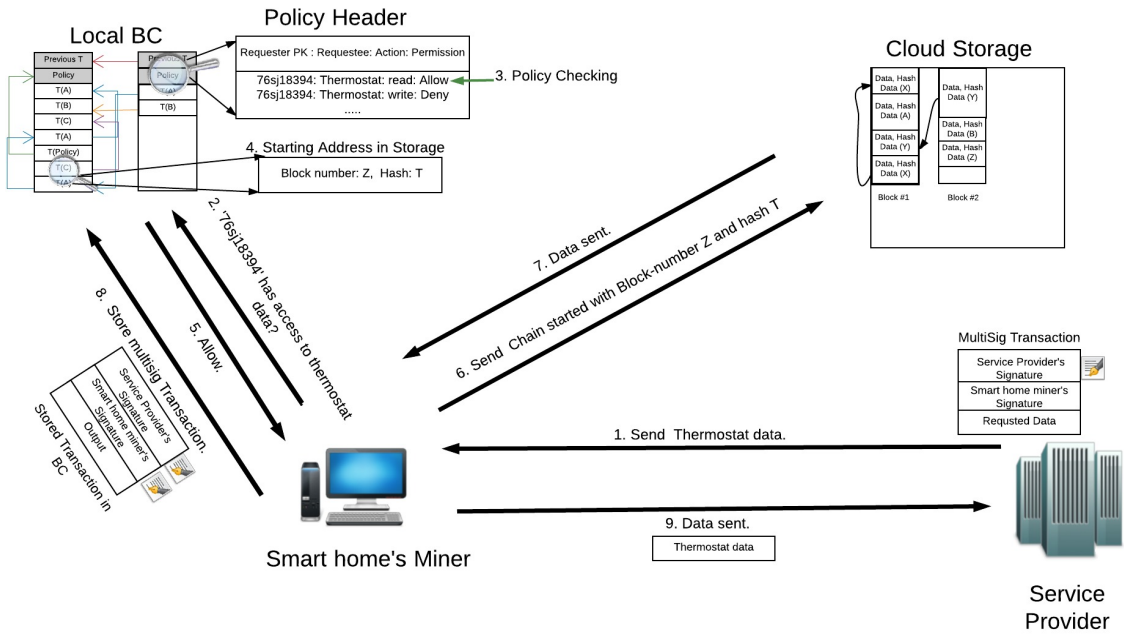

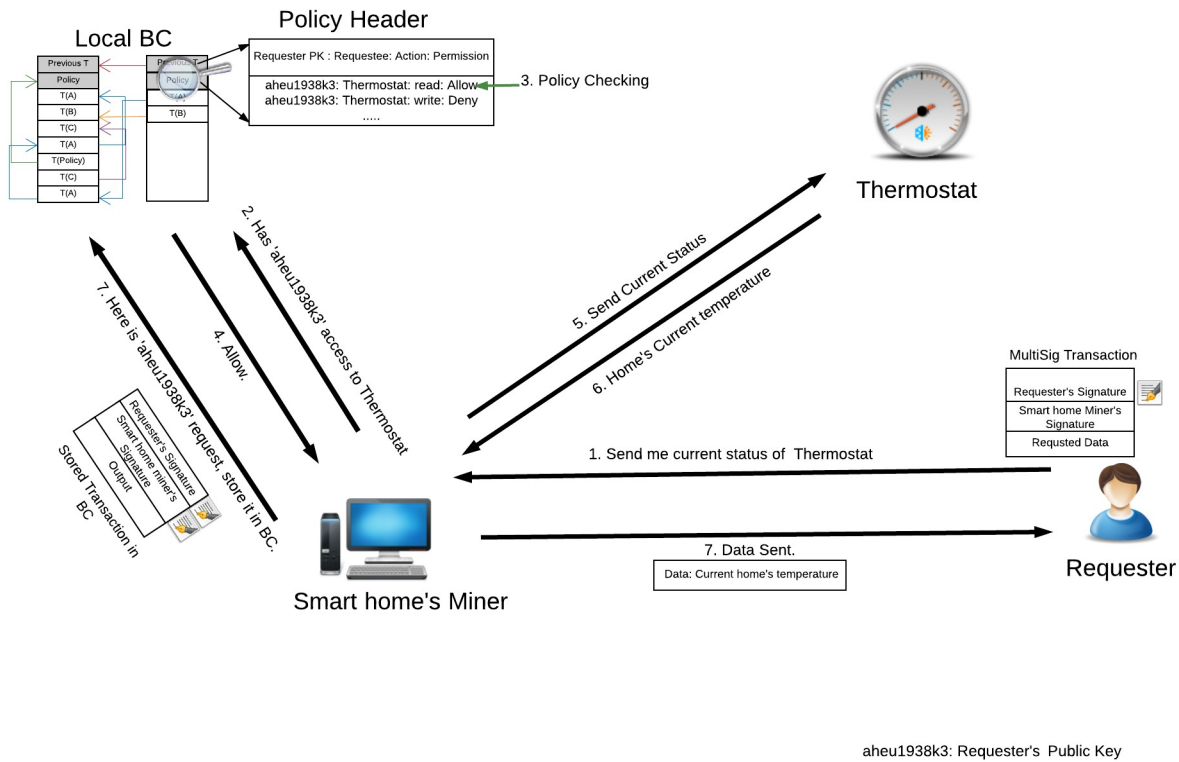

(c)

Fig 2. Process of a) Store Transaction, b) Access Transaction, c) Monitor Transaction.

**Distributed Trust**

We now discuss the mechanisms for ensuring distributed trust in the overlay network. In the overlay network each CH maintains a trust rating for other CHs based on the Beta Reputation System [12] which relies on direct and indirect evidence. In the proposed architecture, CH A has direct evidence about CH B if it verifies a block mined by B. If A receives B's block from CH C, then it has indirect evidence about B. When a CH generates a new block it has to create a multisig transaction which is used for evaluating trust. The CH then sends both the block and multisig transaction to its neighboring CHs. When a new block is received by a CH, it attempts to validate the associated multisig transaction. If it has direct evidence with the block miner or other CHs who signed the multisig transaction, then it randomly verifies a portion of the transactions in the block by checking their signatures. The number of verified transactions by the CH depends on the degree of direct evidence it has about the block miner and its trust assessment of the CHs that provide indirect evidence, where more trustworthy evidence requires fewer random transactions to be checked. If a CH has no direct evidence with the block generator or those who signed it, then it checks all transactions.

As is evident, there are some distinct differences between how BC is employed in Bitcoin and in our proposed architecture. We summarize the key differences in Table 1.

Table 1. Comparison of the Bitcoin BC and the BC employed in different tiers of our proposed architecture.

| # | Studied Parameter | BC in bitcoin | Local BC | Shared BC | Overlay BC |
|---|---|---|---|---|---|
| 1 | BC Visibility | Public | Secure/ Private | Secure/ Private | Public |
| 2 | Transaction chaining | Input / Output | Previous T of the same D | Previous T of the same D in the same H | T are chained to each other/ Output. |
| 3 | Transaction mining | All Ts | All Ts | All Ts | Arbitrary Ts |
| 4 | Mining requirement | Proof of Work | None | None | None |
| 5 | Forking | Not allowed | Allowed | Allowed | Allowed |
| 6 | Double Spending | Prohibited | Not applicable | Not applicable | Not applicable |
| 7 | Transaction verification | Signature | No verification | Signature | Signatures |
| 8 | Transaction parameters | input, output, coins. | Block-number, hash of data, time, output, PK, policy rules. | Block-number, hash of data, time, output, PK, policy rules. | Output, PKs. |
| 9 | Transaction dissemination | Broadcast | Unicast | Unicast | Unicast/ multicast |
| 10 | Deference in block header | puzzle | policies | policies | Not applicable |
| 11 | Blocks stored by miner | All blocks | All blocks | All blocks. | Arbitrary blocks |
| 12 | New block verification | Blocks and Ts in blocks | No verification | No verification | Blocks and Ts in blocks |
| 13 | BC control | No one | Owner | Owner | No one |
| 14 | Miner checks | No one | No one | No one | Other miners and nodes. |
| 15 | Miner trust | Miners are all the same. | Miners are all the same. | Miners are all the same. | Different levels of trust are defined. |
| 16 | How many blocks each T is stored in? | one block | one block | one block | one block |
| 17 | Miner joining overhead | download all blocks in BC. | download all blocks in BC. | Download all blocks in BC. | Download arbitrary blocks in BC |
| 18 | Miner selection | Self-selection | Owner chooses the miner. | Owner chooses the miner | Nodes in cluster choose one node in the cluster as miner. |
| 19 | Miner rewards | Coins | Nothing | Nothing | Not defined |
| 20 | Pool mining | allowed | Cannot be defined. | Cannot be defined. | Cannot be defined. |
| 21 | Malicious miner | Allowed to join | not possible | not possible | Allowed to join |
| 22 | Effects of 51% attack | double spending | not possible. | not possible | Increases the possibility of mining false blocks |
| 23 | Encryption method | Public/ private keys | No need | Public/private keys, shared key | Public/private keys, shared key |

T stands for transaction.
D stands for device.
H stands for home.

## Evaluation

In this section, we evaluate and qualitatively analyze the overhead and performance of the proposed architecture under common security and privacy threats. It is assumed that the adversary (or cooperative adversaries) can be the CH, a device in the home, a node in the overlay network, or the storage. Adversaries are able to sniff communications, discard transactions, create false transactions and blocks, change or delete data in storage, link a user's transactions to each other and sign fake transactions to legitimize colluding nodes. However, they are not able to break the encryption. The main classes of threats are:

- Threat to accessibility: The goal of the adversary is to prevent the legitimate user from getting access to her data or services.
- Threat to anonymity: The goal here is to find the real world identity of the user by analyzing the anonymous transactions and other publicly available information.
- Threats to authentication and access control: The adversary tries to authenticate herself as a legitimate user in order to gain access to data.

We consider the following attacks that threaten accessibility:

- Denial of Service (DOS) Attack: In a DOS attack, the adversary's aim is to prevent the true user from accessing the service or data. In the proposed architecture, an adversary can launch this attack by sending fake transactions or blocks to the overlay network or for particular smart homes. However, the use of requester and requestee PK lists in CHs in our architecture diminishes the effect of this attack. Recall that if neither the multisig transaction requester nor requestee's PK is in these two lists, then the transaction is forwarded to other CHs. Moreover, if a CH receives several unsuccessful access requests from a particular PK, it can block that PK and drop all further requests. However, the adversary can succeed in a DOS attack if it uses different PKs for the attack.
- Modification Attack: To launch this attack, the adversary would have to compromise the cloud storage security. The adversary may then seek to change or delete stored data for a particular user. This user would be able to detect any change in its stored data by comparing hash of the data in cloud with stored hash in its local BC. If a user detects a data breach, she creates a transaction pointing to two transactions: the multisig transaction signed by both the user and the cloud storage containing the true hash of the data and the access transaction signed by both the storage and the user containing invalid hash of the data. The transaction is then sent to a range of CHs who validate the original transactions being referenced. In case of inconsistency in two hashes, the CH informs its nodes of malicious activities by the cloud storage. However, a user cannot recover its data when exposed to this attack.
- Dropping Attack: To launch this attack, the adversary should have control over a CH or a group of CHs. The CHs under the attacker's control should then drop all received transactions and blocks. However, such an attack would be detected since nodes that belong to the constituent clusters would not receive any transactions or service from the network. In our proposed architecture, if such a situation is detected, then all nodes within the same cluster are made aware of it and a new CH is elected.
- Mining Attack: To launch this attack, the adversary must control multiple CHs that work cooperatively and sign the multisig transaction along with a fake block mined by their cooperative. In the proposed trust method all transactions of a received block are validated unless the CH had direct evidence with block miner or those who signed the multisig transaction. In which case, a random portion of blocks are validated based on the trust level. As such, the proposed architecture may not always be able to detect the fake block. It is worth mentioning that even if one CH cannot detect a fake block, other CHs may be able to detect it. As long as one CH can detect a fake block, it can broadcast an alarm and intimate all other CHs.

For breaking anonymity, an adversary tries to link different transactions with different IDs to one real identity in the real world. To avoid this, the proposed architecture allows users to send arbitrary transactions to the overlay network. In addition, IDs and PKs are changeable for each transaction.

The final class of threats is against authentication and access control, where the adversary aims to hack into existing devices in the home. This is detectable by the user since all transactions are mined in the local BC. The other possibility is that the adversary attempts to add a new device to the smart home. This attack is not possible since all devices should be pre-defined by the user and a starting transaction should be mined in the local BC.

The adversary may pose as a SP. Then when it receives the block-number and hash from a user, it can use these two parameters for verifying itself as a true user to storage and manipulate the storage such that it is no longer accessible to the user. In the proposed architecture, each block in the storage can be chained to one further block. Prior to giving the requester these two parameters, the miner stores some data, even an empty block, and points that back to the one which will be given to the requester. By doing this, the requester cannot chain its data to user's data since the given block is already chained.

In summary, the proposed architecture enforces security and privacy properties in each tier through appropriate methods. Table 2 summarizes them in a succinct manner.

Table 2. Entities and methods enforcing security and privacy properties in different tiers.

| Properties | Smart Home | Overlay Network | Cloud Storage |
|---|---|---|---|
| Identity and Authentication | Ledger of transactions | Signatures | Block-number along with Hash |
| Access control | Policy header and transactions in BC | Multisig transaction | Block-number along with Hash |
| Protocol and network security | Encryption | Encryption | Encryption |
| Privacy | Not-private | PK or ID | Block-number along with Hash |
| Trust | Pre-defined | Verification | Signed Hash of data |
| Non-Reputation | Encryption | Signatures | Signed hash of data |
| Policy enforcement | Policy header | PK lists | Accounting |
| Authorization | Policy header and transactions | List of Keys | Accounting |
| Fault tolerance | Medium | High | Low |

Table 3 evaluates scaling performance metrics for mining and all aforementioned transactions in the proposed architecture as a function of key network parameters. For instance, for access transactions, in the best case, when the requester and requestee are in the same cluster, the packet overhead and delay are affected directly by the number of hops between the smart home and the storage. In contrast, computation and memory overhead are constant. Overall, other than new miner joining, actions scale at worst with the number of clusters, which grows more slowly than the number of nodes in the network.

Table 3. Overhead Evaluation.

| | Mining & Trust | $T_{Access}$ | | | $T_{Store}$ | | | $T_{Monitor}$ | | | New miner joining |
|---|---|---|---|---|---|---|---|---|---|---|---|
| | | Best | Average | Worst | Local | Overlay | Cloud | Best | Average | Worst | |
| Packet overhead | O(N) | O (S) | O ((N*S)/2) | O (N*S) | O (1) | O (1) | O (S) | O (S) | O ((N*S)/2) | O (N*S) | O (BS) |
| Delay | O (N/TL) | O (S) | O ((N*S)/2) | O (N*S) | O (1) | O (1) | O (S) | O (S) | O ((N*S)/2) | O (N*S) | O (B*T) |
| Computation overhead | O (N/TL) | O (1) | O (N) | O (N) | O (1) | O (1) | O (1) | O (1) | O (N) | O (N) | O (B*T) |
| Memory overhead | O (BS) | O (1) | O (1) | O (1) | O (1) | O (1) | O (1) | O (1) | O (1) | O (1) | O (BS) |

N: The number of Clusters.
BS: Block Size.
TL: Trust Level.
S: Hops between source and storage.
T: Transactions in each block.
B: Blocks in each BC.

## Conclusion

IoT security and privacy are critical success factors for meeting the high expectations of the technology to transform many aspects of our society and economy. Our proposed blockchain-based IoT architecture handles most security and privacy threats, while considering the resource-constraints of many IoT devices. Our qualitative overhead analysis of the the architecture has shown that it has constant performance overhead at best, and at worst most of its transactions scale with the number of clusters in the network, rather than the number of nodes. While our architecture has been presented in the context of a smart home, it is broadly applicable to most multi-tiered IoT network topologies.

Open questions remain around further reducing vulnerability to denial of service attacks, modification attacks, and the 51% attack for establishing distributed trust. The intrinsic broadcast medium, decentralization, and resource-constraints of IoT are key challenges towards answering these questions. The architecture proposed in this paper lays the groundwork for further research in this area, providing a lightweight, secure and private framework that retains most benefits of blockchain technology.

**Biographies**

Ali Dorri received his bachelor degree in Computer Engineering from Bojnourd University, IRAN, 2012. He then commenced his master degree in Computer Engineering in Islamic Azad University of Mashhad, IRAN, working on Mobile Ad hoc Networks and security issues rising from this sort of network. He now is a Ph.D. candidate in University of New South Wales (UNSW), Sydney. His current research interest covers security and privacy concerns in the context of Internet of Things (IoT), Wireless Sensor Network (WSN) and Vehicular Ad hoc Network (VANET). Moreover, he is working on blockchain and its applications on IoT.

Salil S. Kanhere received his M.S. and Ph.D. degrees, both in Electrical Engineering from Drexel University, Philadelphia. He is currently an Associate Professor in the School of Computer Science and Engineering at the University of New South Wales in Sydney, Australia. His current research interests include Internet of Things, pervasive computing, crowdsourcing, embedded sensor networks, mobile networking, privacy and security. He has published over 150 peer-reviewed articles and delivered over 20 tutorials and keynote talks on these research topics. He is a contributing research staff at Data61, CSIRO and a faculty associate at Institute for Infocomm Research, Singapore. Salil regularly serves on the organising committee of a number of IEEE and ACM international conferences (e.g, IEEE PerCom, ACM MobiSys, ACM SenSys, ACM CoNext, IEEE WoWMoM, IEEE LCN, ACM MSWiM, IEEE DCOSS, IEEE SenseApp, ICDCN, ISSNIP). He currently serves as the Area Editor for Pervasive and Mobile Computing, Computer Communications, International Journal of Ad Hoc and Ubiquitous Computing and Mobile Information Systems. Salil is a Senior Member of both the IEEE and the ACM. He is a recipient of the Humboldt Research Fellowship in 2014.

Raja Jurdak is a Principal Research Scientist at CSIRO, where he leads the Distributed Sensing Systems Group. He has a PhD in Information and Computer Science at University of California, Irvine in 2005, an MS in Computer Networks and Distributed Computing from the Electrical and Computer Engineering Department at UCI (2001), and a BE in Computer and Communications Engineering from the American University of Beirut (2000). His current research interests focus on energy-efficiency and mobility in networks. He has over 100 peer-reviewed journal and conference publications, as well as a book published by Springer in 2007 titled Wireless Ad Hoc and Sensor Networks: A Cross-Layer Design Perspective. He regularly serves on the organizing and technical program committees of international conferences (DCOSS, RTSS, Sensapp, Percomm, EWSN, ICDCS). Dr. Jurdak is an Adjunct Professor at Macquarie University and James Cook University, and Adjunct Associate Professor at the University of Queensland and the University of New South Wales. He is a Senior Member of the IEEE.